# Dirac fermions and flat bands in the ideal kagome metal FeSn

Mingu Kang[1,†], Linda Ye[1,†], Shiang Fang[2], Jhih-Shih You[3], Abe Levitan[1], Minyong Han[1], Jorge I. Facio[3], Chris Jozwiak[4], Aaron Bostwick[4], Eli Rotenberg[4], Mun K. Chan[5], Ross D. McDonald[5], David Graf[6], Konstantine Kaznatcheev[7], Elio Vescovo[7], David C. Bell[8,9], Efthimios Kaxiras[2,8], Jeroen van den Brink[3], Manuel Richter[3,10], Madhav Prasad Ghimire[3,11], Joseph G. Checkelsky[1,\*] & Riccardo Comin[1,◊]

[1]Department of Physics, Massachusetts Institute of Technology, Cambridge, Massachusetts 02139, USA.

[2]Department of Physics, Harvard University, Cambridge, Massachusetts 02138, USA.

[3]Leibniz Institute for Solid State and Materials Research, IFW Dresden, Helmholtzstr. 20, 01069 Dresden, Germany.

[4]Advanced Light Source, E. O. Lawrence Berkeley National Laboratory, Berkeley, California 94720, USA.

[5]National High Magnetic Field Laboratory, Los Alamos National Laboratory, Los Alamos, New Mexico 87545, USA.

[6]National High Magnetic Field Laboratory, Tallahassee, Florida 32310, USA.

[7]National Synchrotron Light Source II, Brookhaven National Laboratory, Upton, New York 11973, USA.

[8]John A. Paulson School of Engineering and Applied Sciences, Harvard University, Cambridge, Massachusetts 02138, USA.

[9]Center for Nanoscale systems, Harvard University, Cambridge, Massachusetts 02138, USA.

[10]Dresden Center for Computational Materials Science (DCMS), TU Dresden, Dresden 01062, Germany.

[11]Central Department of Physics, Tribhuvan University, Kirtipur, 44613, Kathmandu, Nepal.

[†]These authors have contributed equally to this work;

[\*]checkelsky@mit.edu; [◊]rcomin@mit.edu

**The kagome lattice based on 3*d* transition metals is a versatile platform for novel topological phases hosting symmetry-protected electronic excitations and exotic magnetic ground states. However, the paradigmatic states of the idealized two-dimensional (2D) kagome lattice – Dirac fermions and topological flat bands – have not been simultaneously observed, partly owing to the complex stacking structure of the kagome compounds studied to date. Here, we take the approach of examining FeSn, an antiferromagnetic single-layer kagome metal with spatially-decoupled kagome planes. Using polarization- and termination-dependent angle-resolved photoemission spectroscopy (ARPES), we detect the momentum-space signatures of coexisting flat bands and Dirac fermions in the vicinity of the Fermi energy. Intriguingly, when complemented with bulk-sensitive de Haas-van Alphen (dHvA) measurements, our data reveal an even richer electronic structure that exhibits robust surface Dirac fermions on specific crystalline terminations. Through band structure calculations and matrix element simulations, we demonstrate that the bulk Dirac bands arise from in-plane localized Fe-3*d* orbitals under kagome symmetry, while the surface state realizes a rare example of fully spin-polarized 2D Dirac fermions when combined with spin-layer locking in FeSn. These results highlight FeSn as a prototypical host for the emergent excitations of the kagome lattice. The prospect to harness these excitations for novel topological phases and spintronic devices is a frontier of great promise at the confluence of topology, magnetism, and strongly-correlated electron physics.**

The kagome lattice is a two-dimensional network of corner-sharing triangles (Fig. 1a), which has originally gained the spotlight as a platform for frustration-driven exotic spin-liquid phases.[1,2] Recent theoretical investigations have focused on the emergent electronic excitations engendered by the special geometry of the kagome network, whose unique combination of lattice symmetry, spin-orbit coupling, and unusual magnetism sets an ideal stage for novel topological phases.[3–8] Viewed as an isolated layer, the kagome lattice hosts a flat band and a pair of Dirac bands, the latter protected by symmetry, in analogy to graphene (Fig. 1b).[3,4] Compounded with spin-orbit coupling and a net magnetization, the two-dimensional kagome lattice realizes a 2D Chern insulator phase with quantized anomalous Hall conductance at 1/3 and 2/3 fillings.[5] When these quantum anomalous Hall layers are stacked along the third dimension, the interlayer interaction drives the mass gap to be closed and reopened along the stacking axis, transforming the system into a three-dimensional Weyl semimetal phase with broken time-reversal symmetry.[7,9] At the same time, a flat band on the kagome lattice also carries a finite Chern number, and mimics the phenomenology of Landau levels, without an external magnetic field.[8,10] As a result, the fractional quantum Hall state can be realized at a partial filling of these flat bands, further enriching the spectrum of topological phases that can be harnessed within the kagome lattice.

These promising theoretical proposals have driven and guided recent experimental efforts toward the realization and study of topological kagome metals based on binary and ternary intermetallic compounds[11–22]. At variance with other widely studied $s$ or $p$ orbital-based topological systems that are close to the non-interacting limit, the kagome lattice in these intermetallic materials is populated by the low-energy $3d$ electrons of transition metals (Fig. 1a), and thus provide an ideal platform to study the interplay of electronic topology and strong correlations. Correspondingly, not only topological electronic structures but also rich intrinsic magnetism can be found in the $3d$ kagome metal series. The combination of these two aspects gives rise to intrinsic anomalous Hall conductivity via various mechanisms.[11,13,15,19,20]

Despite the great potential and rich phenomenology of this family of materials, the experimental realization of the electronic structure of an idealized 2D kagome lattice, namely the Dirac fermions and topological flat bands (Fig. 1b), in bulk magnetic kagome crystals has remained an outstanding challenge. For instance, in the binary intermetallic $T_mX_n$ kagome series ($T$ = Mn, Fe, Co, $X$ = Sn, Ge, $m$:$n$=3:1, 3:2, 1:1) with various stacking sequences of kagome and spacer $X$ layers (Fig. 1c-e), the quasi-2D Dirac electronic structure has been detected only in

Fe$_3$Sn$_2$[15] but not in Mn$_3$Sn.[13] Rather, Mn$_3$Sn and ternary kagome compound Co$_3$Sn$_2$S$_2$, three-dimensional magnetic Weyl points have been identified as the potential origin for the chiral anomaly in transport[13,19], as also confirmed by band structure calculations[7,19–21]. For what concerns the flat bands, a diverging density of states was found in Fe$_3$Sn$_2$ and Co$_3$Sn$_2$S$_2$ using scanning tunneling spectroscopy, however, no signatures of these nondispersive excitations in momentum space have been reported so far, presumably due to the complex stacking structures (Fig. 1d).[18,22] Given the premises, the investigation of a material with a simpler stacking structure is desired to provide a prototypical realization of the electronic structure of the kagome lattice.

In this study, we report the electronic structure of single-layered kagome metal FeSn, unique among its $T_mX_n$ sibling compounds in that it is based on isolated and spatially decoupled kagome planes (Fig. 1e, see Methods and Supplementary Fig. 1 for details on synthesis and characterizations). Compared to the previously-studied Mn$_3$Sn and Fe$_3$Sn$_2$ structures (Fig. 1c,d), FeSn is the one closest to the 2D limit, setting an ideal stage to investigate the electronic structure of the 2D kagome network in a bulk crystal. FeSn is magnetically ordered, with Fe moments ferromagnetically aligned within each kagome plane but antiferromagnetically coupled along the c-axis.[23] This magnetic state allows for a simple hopping model, free from the complications of a non-collinear magnetic texture as instead is found in Mn$_3$Sn-type kagome antiferromagnets.[11] As detailed below, our comprehensive study of the electronic structure of FeSn – combining ARPES, magneto-quantum oscillations, and density functional theory (DFT) calculations – reveals the rich phenomenology of the kagome lattice in its full variety, featuring the coexistence of bulk and surface Dirac fermions, as well as the long sought-after flat bands.

In Figure 2, we summarize our photoemission experiments on FeSn. We first note that FeSn can expose two possible surface terminations upon cleaving, namely the kagome and Sn terminations marked as A and X in Fig. 1e. We determined that the surface termination can be uniquely identified *in situ* using X-ray photoelectron spectroscopy (XPS) on Sn 4*d* core levels as shown in Fig. 2a,b (see Methods and Supplementary Fig. S2,3 for detailed analysis of XPS spectra and atomic force microscopy characterization of the surface terminations), and measured valence band structure on both terminations. Interestingly, our ARPES data uncover a strong dependence of the valence band structure on surface termination. In Fig. 2d,f, we first present the Fermi surface (d) and energy-momentum dispersion (f) from the kagome termination. The most prominent feature on the Fermi surface is a circular electron pocket (dashed circle in Fig. 2d) centered at the corner of the hexagonal Brillouin zone (K-point), which arises out of the Dirac bands expected from the kagome tight binding model (Fig. 1b) and previously observed in $Fe_3Sn_2$.[15] The energy-momentum dispersion across the K-point (Fig. 2f) confirms the presence of a clear Dirac cone structure (DC1) with linear crossing at $E_{D1} = -0.43 \pm 0.02$ eV. To closely visualize the momentum-space structure of the Dirac bands, we show a series of constant energy maps in Fig. 2g-j, measured at +0.4 eV, +0.2 eV, 0 eV, and –0.2 eV with respect to $E_{D1}$. Far above $E_{D1}$, the large electron pocket centered at the K-point (Fig. 2g) can be seen, which shrinks to a single Dirac point (Fig. 2i) and reopens (Fig. 2j) linearly as the energy crosses $E_{D1}$. The velocity of Dirac fermions in FeSn is $(1.7 \pm 0.2) \times 10^5$ m/s, an order of magnitude lower than that of graphene and in close range of that of $Fe_3Sn_2$[15] and Fe-based superconductors[24–26], possibly reflecting the more correlated nature of Fe-3$d$ electrons. Overall, our ARPES data directly establish the Dirac fermiology of kagome-derived bands in FeSn.

Intriguingly, the electronic structure measured on the Sn termination is even richer than the kagome termination, as shown in Fig. 2c,e. The Fermi surface (Fig. 2c) exhibits a triangular electron pocket centered at the K-point in addition to the circular pocket from DC1. The band dispersion shown in Fig. 2e reveals that the new pocket arises from a second Dirac cone (DC2) with crossing at $E_{D2} = -0.31 \pm 0.02$ eV. The binding energy and dispersion of DC1 is unaltered on this termination. Further, DC1 and DC2 exhibit very different trigonal warping patterns away from the Dirac point (black dashed circle and triangle in Fig. 2c), despite their similar Dirac velocity. This aspect hints at the different origin of two Dirac bands, and rules out other scenarios, such as spin-splitting, layer-splitting, quantum-well states[27] or bosonic shake-off replicas.[28] Instead, the inequivalence of the electronic spectra from the two terminations provides a direct insight on the nature of DC1 and DC2: the former is a bulk Dirac band that manifests itself independently of the surface termination, while the latter represents a surface Dirac state that is observed only on the Sn termination. Photon energy-dependent ARPES measurements further reveal that the dispersions of DC1 and DC2 are unaltered along the out-of-plane direction (Supplementary Fig. S4), reflecting the 2D nature of Dirac fermions in FeSn.

The surface vs. bulk origin of the two Dirac bands can be further pinned down using a bulk sensitive probe of the electronic structure: here we focus on the de Haas-van Alphen effect, which as a thermodynamic quantity exclusively measures the quantized Landau level formation of *bulk* Fermi surfaces. Using torque magnetometry at high magnetic fields and low temperatures

(Methods), we resolve dHvA oscillations (see Supplementary Fig. S5) with frequencies summarized in Fig. 3a as circles. Multiple frequencies are observed which vary systematically as a function of field orientation with respect to the kagome plane normal ($\theta$, see inset of Fig. 3a). Magnetoresistance Shubnikov-de Haas (SdH) oscillations are also observed (see Supplementary Fig. S6) whose frequencies comprises a subset of the dHvA frequencies as marked as triangles in Fig. 3a. We index these branches as $\alpha_{1,2,3}$, $\beta_{1,2}$, $\gamma_{1,2}$ and $\delta$ based on their qualitative evolution with $\theta$. Most importantly, the frequencies of the $\beta$, $\gamma$ and $\delta$ bands remain finite during a complete $\theta$ rotation, while the $\alpha$ band frequency diverges as $1/\cos\theta$ as the magnetic field is tilted toward the kagome plane. The former behavior is a characteristic of three-dimensional closed Fermi pockets, while the latter is indicative of quasi-2D Fermi sheets (for an ideal 2D Fermi surface, a $f(\theta) = f_0 / \cos\theta$ dependence is expected where $f_0$ is proportional to the area of Fermi surface $A_F$). The values of $f_0$ ($A_F$) extracted from the fit (dashed lines in Fig. 3a) are 1310 T (0.125 Å$^{-2}$), 3642 T (0.348 Å$^{-2}$), and 6755 T (0.656 Å$^{-2}$) for $\alpha_1$, $\alpha_2$, and $\alpha_3$ respectively. To further characterize these Fermi surfaces, we show in Fig. 3b the damping of the quantum oscillation amplitudes with elevated temperature fitted with a Lifshitz-Kosevich formula (see Methods). The obtained effective masses $m^*$ of the $\alpha_1$, $\alpha_2$, and $\alpha_3$ bands are (5.4 ± 0.4) $m_e$, (3.1 ± 0.2) $m_e$, and, (4.3 ± 0.3) $m_e$ respectively.

To validate the correspondence between the observations from ARPES and quantum oscillations, we compare the experimental parameters of DC1, DC2, $\alpha_1$, $\alpha_2$, and $\alpha_3$ in Table 1. In magneto-quantum oscillations, $v_F$ can be obtained from $m^*$ by assuming a Dirac dispersion[17], i.e. $v_F = \sqrt{E_F/m^*}$. An excellent agreement in both $A_F$ and $v_F$ is obtained between DC1 and $\alpha_2$, suggesting that they represent the same band. This equivalence confirms the bulk origin of DC1, whose Landau orbit under high magnetic field appears as $\alpha_2$ in both dHvA and SdH experiments. In contrast, both $A_F$ and $v_F$ of DC2 markedly deviate from those of $\alpha_1$ and $\alpha_3$, by a factor of almost 2. Instead, a close comparison with calculated dHvA spectrum (Supplementary Fig. S7) suggests that $\alpha_1$ and $\alpha_3$ originate from different quasi-2D Fermi surfaces centered at $\Gamma$. The absence of a bulk band corresponding to DC2 thus demonstrates its surface origin as inferred from the termination-dependent ARPES spectra above. In sum, by combining complementary photoemission and quantum oscillations experiments, we confirmed the quasi-2D Dirac fermiology of FeSn, and also revealed the unusual coexistence of surface and bulk Dirac fermions in the single compound.

To understand the origin of surface and bulk Dirac fermions, we extended the tight binding calculations of the ideal two-dimensional kagome lattice (Fig. 1b) to incorporate the *d*-orbital degrees of freedom. Further, we performed DFT calculations of FeSn in both bulk and slab geometries. In the single-layer kagome tight binding model with a *d*-orbital basis, five separate Dirac points emerge at the K-point from the five 3*d* orbitals (Supplementary Fig. S8). Complementary bulk DFT calculations (Supplementary Fig. S9) reveal that the Dirac points with different orbital character respond very differently to the interlayer coupling: the Dirac points with in-plane orbital character ($d_{xy}$ and $d_{x2-y2}$) retain their 2D nature and are unaffected when embedded in the bulk Brillouin zone, while those with out-of-plane orbital character ($d_{xz}$, $d_{yz}$, and $d_{3z2-r2}$) acquire a pronounced $k_z$ dispersion and lose their characteristic 2D kagome features in the bulk model. Accordingly, near the Fermi level our bulk DFT predicts a single 2D Dirac band with $d_{xy}$ + $d_{x2-y2}$ characters at E ≈ – 0.4 eV (Fig. 4a and Supplementary Fig. S9), which is in close agreement with the experimentally-observed bulk Dirac cone (DC1). The robustness of the 2D Dirac dispersion in the bulk kagome lattice is in stark contrast to the case of graphene, where the Dirac cone with $p_z$ orbital character is strongly susceptible to interlayer interactions and loses its characteristic linear dispersion or 2D nature in multilayer or bulk form.[29,30] These findings suggest that the careful engineering of localized 3*d*-orbital character is a key to realize the desired kagome electronic bands in bulk magnetic kagome crystals.

The surface Dirac band (DC2) is also reproduced in DFT calculations based on a slab geometry. Figure 3c,d display the band structure of FeSn slabs composed of eight kagome layers terminated with kagome layer on one side and Sn layer on the other side. (schematically shown in the insets of Fig. 3c,d). In this model, the six inner kagome layers mimic the bulk local environment (marked as 'bulk', orange circles), while the outer kagome layers are subject to the surface potential of each termination (marked as 'surf', red and blue circles). For simplicity, in Fig. 3c,d we only weight second outermost kagome layers as a 'bulk' state, since all six inner kagome layers are essentially degenerate in terms of Dirac bands (see Supplementary Fig. S10). First, one finds a termination-independent Dirac cone (orange) arising from the 'bulk' kagome layers at E ≈ –0.4 eV, which corresponds to DC1 in bulk DFT and ARPES. At the same time, an additional Dirac band (red) localized within the surface kagome layer emerges at E ≈ –0.3 eV in the Sn-terminated slab, which closely reproduces the DC2 band observed by ARPES (see Fig. 2e). The orbital analysis presented in Supplementary Fig. S10 reveals that the surface Dirac state (DC2) possesses an identical orbital character ($d_{xy} + d_{x2-y2}$) to the bulk Dirac state (DC1), indicating that DC2 is a surface resonant state emerging from DC1 under the surface potential. Intriguingly, the surface state is strongly localized on the topmost kagome layer, wherein the spins are ferromagnetically aligned under the intrinsic 'A-type' antiferromagnetism of FeSn. Therefore, the surface state (DC2) of FeSn realizes a rare example of fully spin-polarized 2D Dirac fermions (Fig. 3c), which far surpasses, for example, the partial (≈ 25 %) spin-polarization in the graphene/ferromagnet heterostructures,[31] and is highly desirable for realizing fast-switching/low-power spintronic devices, spin-superconductors,[32] and high-temperature quantum anomalous Hall effect.

The simple single-layer structure of FeSn also enables a detailed analysis of the photoemission intensity pattern of the kagome-derived Dirac cone, which conveys the phase information of the initial state wavefunction. Such analysis has been previously utilized to uncover helical spin textures and chirality of Dirac fermions in three-dimensional topological insulators and graphene,[33–35] but has never been applied to a kagome lattice. As shown in Fig. 2k,l, the photoemission intensity is strongly modulated around the Dirac cone, a direct consequence of phase interference between wave functions from different kagome sublattices. The intensity modulation follows a $\cos\phi$ function (where $\phi$ is an azimuthal rotation angle around the Dirac cone), with both maximum and minimum along the G-K direction but at opposite momenta above (Fig. 2k) and below (Fig. 2l) the Dirac point, identical to the case of graphene.[34,35] As shown in Fig. 2n,o, our simulation based on sublattice interference of kagome initial state wavefunctions with Berry phase π (Fig. 2m) closely reproduces this intensity pattern, demonstrating the chirality of kagome-derived Dirac fermions in FeSn (See Supplementary Informations for details).

Having fully characterized the kagome-derived Dirac states, we also searched for the topological flat bands in FeSn, the other defining feature of an ideal kagome lattices. Despite surging theoretical interests on the physics of flat bands,[8,36,37] their direct signatures have been elusive: for example, flat bands reported in other kagome systems including $Fe_3Sn_2$ and $Co_3Sn_2S_2$ were confined to subregions of the Brillouin zone possibly due to the complex interlayer interactions in these systems.[18,22] Based on our orbital analysis above, we could infer that FeSn might be an ideal system where flat bands constructed from in-plane $d$-orbitals are invulnerable to interlayer interactions and retain their nondispersive character in all three momentum-space directions. Accordingly, our bulk DFT calculation reveals quasi-2D nearly flat bands (with the bandwidth about 1/5 of that of the Dirac bands) with $d_{xy}$ and $d_{x2-y2}$ characters at about 0.5 eV above

Fermi level (see Supplementary Fig. S9 and Fig. 4a). In combination with the observed Dirac cone structure, the complete prototypical kagome band structure of Fig. 1b could be thus realized or mimicked in FeSn using electronic bands derived from in-plane 3*d* orbitals.

The aforementioned quasi-2D nearly flat band cannot be directly observed by ARPES as it lies above the Fermi level. Instead, we can search for a signature of flat bands from other *d*-orbital degree of freedom, which can arise at various energies (see Supplementary Fig. S8 for *d*-orbital-based kagome tight-binding model). Fig. 4b,c display the experimental band structures of FeSn along Γ-K-M high symmetry directions measured with linear horizontal (LH) and linear vertical (LV) polarizations respectively. The signature of flat bands is absent under LH polarization, while LV polarization reveals a strikingly nondispersive excitation near the Fermi level ($E_{flat}$ = –0.23 ± 0.05 eV). Accordingly, as shown in the constant energy map at $E_{flat}$ in Fig. 4d, the spectral weight of this band is uniformly distributed across almost the full Brillouin zone except around the K-points due to the intensity leakage from the Dirac bands. This observation represents the first momentum-space evidence of the flat band in the kagome system. The experimental bandwidth of the flat band is less than 1/10 of that of the Dirac bands and comparable to what has been observed in f-electron systems.[38] Unlike the latter cases, however, the kagome-derived flat band arises from a destructive phase interference of hopping in a frustrated geometry (inset of Fig. 4c), and is thus intrinsically topological with finite spin-orbit coupling.[10] In real-space, this phase interference effectively localizes the wavefunction into a single hexagon as depicted in the inset of Fig. 4c. Such localization to the subregion of real space is similar to the case of engineered flat band in the magic-angle twisted bilayer graphene, with electron localized to the AA-stacked region of the Moiré superlattice.[39] Comparing the length scale of localization *d*, the kagome lattice (in ideal case) evidently promotes a stronger localization ($d \approx 5$ Å) than the Moiré superlattice ($d \approx 50$ Å), which directly implies stronger effective Coulomb energy scale by $U = e^2/4\pi\varepsilon d$. The identification of the flat band opens up important opportunities for engineering new correlated electron phases in the kagome lattice via local electrostatic gates (tuning the flat bands near the Fermi level) and by controlling the strength of magnetic exchange splitting. Altogether, our discovery and extensive analysis of Dirac cone and flat bands in the ideal kagome metal FeSn unlock new perspectives and avenues for the realization of novel correlated topological phases and spintronic devices based on kagome lattices.

**Methods**

**Sample growth and characterizations.** Single crystals of FeSn were grown using a chemical vapor transport technique with $I_2$ as a transport agent. Fe powder (Alfa Aesar, 99.998%) and Sn powder (Sigma Aldrich, 99.99%) were loaded in a quartz tube with ~3 mg/cm$^3$ $I_2$. The evacuated quartz tube was put in a temperature gradient of 520 °C (source) - 680 °C (sink) in a horizontal three zone furnace. Thin plate-like, hexagonal single crystals were obtained and a typical growth duration lasts from 3 weeks to 1 month. The phase of the grown crystals was confirmed with powder X-ray diffraction. Basic transport properties were measured with a standard five-probe configuration in a commercial cryostat.

**Angle-resolved photoemission spectroscopy experiments.** ARPES experiments were performed at two different synchrotron beamlines: the main data were acquired at Beamline 7 (MAESTRO) of the Advanced Light Source, and preliminary experiments were conducted at beamline 21-ID-1 (ESM-ARPES) of the National Synchrotron Light Source II. The two ARPES endstations are respectively equipped with R4000 and DA30 hemispherical electron analyzer (Scienta Omicron). FeSn samples were cleaved inside ultrahigh vacuum chamber with a base pressure better than $4 \times 10^{-11}$ torr. The ARPES data were acquired within 6 hours after cleaving to minimize the effect of surface degradation. All datasets were collected at 20 K, except the one in Supplementary Fig. S4a which was collected at 80 K. The lateral size of the beam was smaller than $20 \times 10$ μm$^2$. Fermi surfaces and energy-momentum dispersions presented in Fig. 2, Supplementary Fig. S11, and Supplementary Fig. S13 were acquired with 92 eV and 140 eV photons which maximize the visibility of Dirac bands. We mainly used LH polarized photons unless otherwise specified. The energy and momentum resolutions were better than 20 meV and 0.01 Å$^{-1}$ respectively. Photon energy dependence in Supplementary Fig. S4 were scanned from 80 eV to 150 eV which covers the complete Brillouin zone of FeSn in $k_z$ direction.

**X-ray photoelectron spectroscopy.** XPS experiments were conducted at Beamline 7 (MAESTRO) of Advanced Light Source using R4000 hemispherical electron analyzer (Scienta Omicron). XPS spectra were measured on the same *in situ* cleaved samples where ARPES experiments were conducted. Before acquiring XPS spectra, we optimized beam position to a large single domain by monitoring the clarity of ARPES spectra. The XPS experimental geometry was such that the analyzer was placed normal to the sample surface while the beam comes from 55° with respect to the sample normal. We acquired XPS spectra from 7 different FeSn samples: four of them represent Sn termination, while the other three represent kagome termination (See Supplementary Fig. S2 for the full dataset). For comparison, we also acquired XPS spectra of $Fe_3Sn_2$ on the same condition (see Supplementary Fig. S2 for detailed comparison).

**Magneto-quantum oscillations.** The magneto-quantum oscillation experiments were performed at the National High Magnetic Field Laboratory (NHMFL). Temperature and angular dependence of the oscillations were examined to reveal the effective mass and dimensionalities of the Fermi surfaces of interest.

The de Haas-van Alphen (dHvA) effects in the magnetic torque were measured using piezoresistive cantilevers (Seiko PRC-400 at the DC field facility and Seiko PRC-150 at the pulsed field facility) under $^3$He or $^4$He atmosphere. No signature of a magnetic phase transition is observed up to 65 T and the transverse magnetization perpendicular to the applied field is estimated to be less than 0.0025 $\mu_B$ per Fe, implying a minimal change in the magnetic structure at high fields where oscillations were observed. We performed additional in-house torque measurements in a superconducting magnet using both capacitive (Cu:Be foil, 10-25 μm) and piezoresistive (SCL Sensortech PRSA-L300) cantilevers.

The Shubnikov-de Haas (SdH) oscillations are observed in magnetoresistance on a thin piece of crystal (~ 6 μm thick) structured with focused ion beam (FIB) and the measurements were performed in DC fields up to 35 T with $^3$He atmosphere. The FIB fabrication was performed with a FEI Helios Nanolab 600 dual beam microscope with a Ga ion beam flux of 21 nA at the magnification of 350.

The oscillatory patterns of both dHvA and SdH oscillations were analyzed with Fast Fourier transform (FFT) as a function of inverse fields. Each set of oscillatory amplitude is modulated by both the thermal and Dingle (residual impurity scattering) damping factors $R_T^i R_D^i$. Here $i$ labels the $i$-th band, and the thermal damping factor is given by $R_T^i = \frac{2\pi^2 k_B T m_i^*}{\hbar e B} \sinh^{-1}\left(\frac{2\pi^2 k_B T m_i^*}{\hbar e B}\right)$ and the Dingle damping factor is given by $R_D^i = \exp\left[-\frac{2\pi^2 k_B T_D^i m_i^*}{\hbar e B}\right]$. $\hbar, k_B$ are the reduced Planck constant and the Boltzmann constant, respectively. Fitting the temperature/field dependence of the oscillation (FFT) amplitudes yields the effective mass $m^*$ and Dingle temperature $T_D$. Here $B$ is taken as the average field of the FFT window from $B^{-1} = \frac{1}{2}(B_{min}^{-1} + B_{max}^{-1})$.

**Bulk and surface electronic structure calculation from the first principles.** Density functional theory (DFT) calculations were performed with the full-potential local-orbital (FPLO) code,[40] version 18.00. The exchange and correlation energy was considered in the generalized gradient approximation (GGA) using the parameterization of Perdew, Burke, and Ernzerhof (PBE-96).[41] Self-consistent calculations were carried out using the four-component fully relativistic mode of FPLO. The following basis states were treated as valence states: Fe: 3s, 3p, 4s, 5s, 3d, 4d, 4p and Sn: 4s, 4p, 4d, 5s, 6s, 5d, 5p, 6p. We used a linear tetrahedron method with 12 × 12 × 12 subdivisions in the full Brillouin zone for the bulk and 8 × 8 × 1 subdivisions for the slabs. We used the experimental structural data. Using the PYFPLO module of the FPLO package,[40] we built a tight-binding Hamiltonian by projecting the Bloch states onto atomic-orbital like Wannier functions associated with Fe 3d and 4s states, and Sn 5s and 5p states. We used this Hamiltonian to compute the Fermi surface, the de Haas van Alphen spectrum and the $k_z$-integrated spectrum shown in Fig. 4. For the Wannier construction we used a mesh of 8 × 8 × 8 subdivisions in the full Brillouin zone.

In order to simulate the [001] surface state, slabs of various thickness, ranging from four atomic layers (one antiferromagnetic unit cell with two kagome layers and two Sn layers) to 16 atomic layers (four antiferromagnetic unit cells with 8 kagome layers and 8 Sn layers). In all cases, we fixed a vacuum of 1.7924 nm, which is four times the lattice parameter $c$ and kept the atomic distances and bond angles as in the bulk. The slabs are terminated with kagome layer on one side and Sn layer on the other side. Thus, they are stoichiometric and their electronic structure can be projected to either termination.

We also performed total energy calculations to estimate the cleavage energy using a k-mesh with 24 × 24 × 1 subdivisions. We find convergence of the cleavage energy with layer thickness already for a slab with eight atomic layers. The calculated GGA cleavage energy amounts to 2.0 J/m² which is significantly larger than that of graphite (0.4 J/m²) and smaller than that of the isotropic three-dimensional (3D) metallic compound FeAl (6 J/m²). We conclude that, from a chemical point of view, FeSn is not a 2D system but may be called an anisotropic 3D system. This conclusion is supported by similar band dispersions within the x-y plane (Γ-M-K-Γ) and perpendicular to that plane (Γ-A, L-M, K-H) as shown in Supplementary Fig. S9.

**Acknowledgements**

We are grateful to C. Felser, L. Levitov, and A. Fahimniya for fruitful discussions. M.P.G., J.-S.Y., J.I.F., S.F., M.R., and J.v.d.B. thank Ulrike Nitzsche for technical assistance in maintaining computing resources at IFW Dresden. R.C. acknowledges support from the Alfred P. Sloan Foundation. This research was funded, in part, by the Gordon and Betty Moore Foundation EPiQS Initiative, Grant No. GBMF3848 to J.G.C. and ARO Grant No. W911NF-16-1-0034. M.K., L.Y., S.F., and M.P.G. acknowledge support by the STC Center for Integrated Quantum Materials, NSF grant number DMR-1231319. M.K. acknowledges support from the Samsung Scholarship from the Samsung Foundation of Culture. L.Y. acknowledges support from the Tsinghua Education Foundation. The computations in this paper were run on the ITF/IFW computer clusters (Dresden, Germany) and Odyssey cluster supported by the FAS Division of Science, Research Computing Group at Harvard University.  M.R and J.v.d.B. acknowledge support from the German Research Foundation (DFG) via SFB 1143, project A5. M.P.G. thanks the Alexander von Humboldt Foundation for financial support through the Georg Forster Research Fellowship Program, Germany. J.-S.Y. and J.I.F. thank the IFW excellence programme. Operation of ESM beamline at the National Synchrotron Light Source is supported by U.S. Department of Energy (DOE) Office of Science User Facility Program operated for the DOE Office of Science by Brookhaven National Laboratory under Contract No. DE-AC02-98CH10886. D.C.B. acknowledges use of the Center for Nanoscale Systems (CNS), a member of the National Nanotechnology Coordinated Infrastructure Network (NNCI), which is supported by the National Science Foundation under NSF award no. 1541959.

**Author contributions**

M.K. performed the ARPES experiment and analyzed the resulting data with assistance from A.L., C.J., A.B., E.R., K.K., and E.V. L.Y. synthesized and characterized the single crystals and performed the quantum oscillation experiments with assistance from M.K.C., R.D.M., and D.G. M.P.G. and S.F. performed the theoretical calculations with assistance from J.-S.Y., J.I.F. J.v.d.V., M.R., and E.K. M.H. and M.K. conducted the AFM measurements. D.C.B. performed the electron microscopy study. All authors contributed to writing the manuscript. J.G.C. and R.C. supervised the project.

**Competing financial interests**

The authors declare no competing financial interests.

**Data availability**

The data that support the plots within this paper and other findings of this study are available from the corresponding author upon reasonable request.

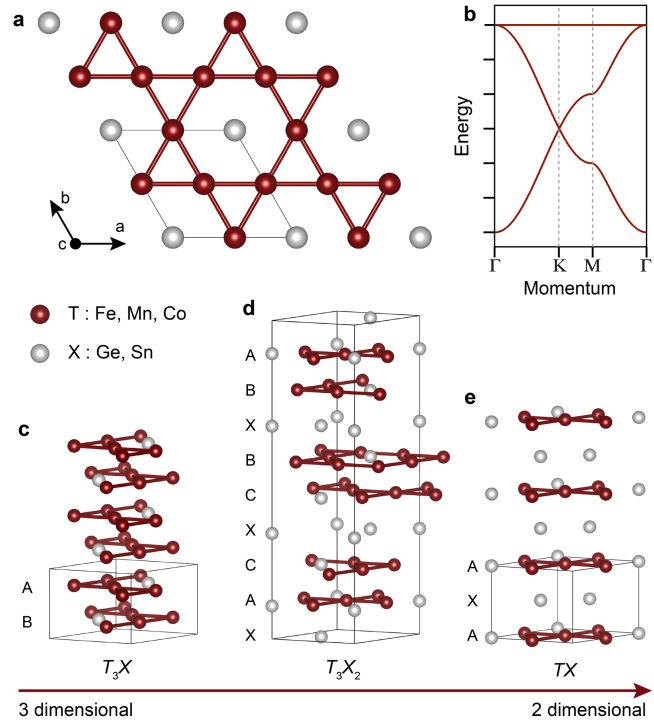

**Figure 1 | Crystal structure of binary kagome metals. a,** Top view of the kagome plane in binary kagome metals $T_mX_n$. The kagome network consists of $3d$ transition metal atoms (Tm: Fe, Mn, Co) with space-filling X: Sn, Ge atoms at the center of hexagon. The in-plane unit cell is marked with the parallelogram. **b,** Tight-binding band structure of kagome lattice exhibiting two Dirac bands at the K-point and a flat band across the whole Brillouin zone. **c-e,** Stacking sequences of the binary kagome metal series $T_mX_n$ with m:n = 3:1, 3:2, and 1:1 respectively. Structural unit cells are marked with solid line. The kagome layers labeled with A-C have different in-plane lattice offset. Spacing layers consisting of X atoms in hexagonal arrangement are labeled with X. The structural two-dimensionality increases with increasing the ratio of X to T. In the $TX$ (1:1) structure (e), the kagome layers are perfectly aligned with one another and are interleaved with X layers, while in the $T_3X$ structure (d), neighboring kagome layers are shifted with respect to each other. In the $T_3X_2$ structure (c) both types of stacking coexist.

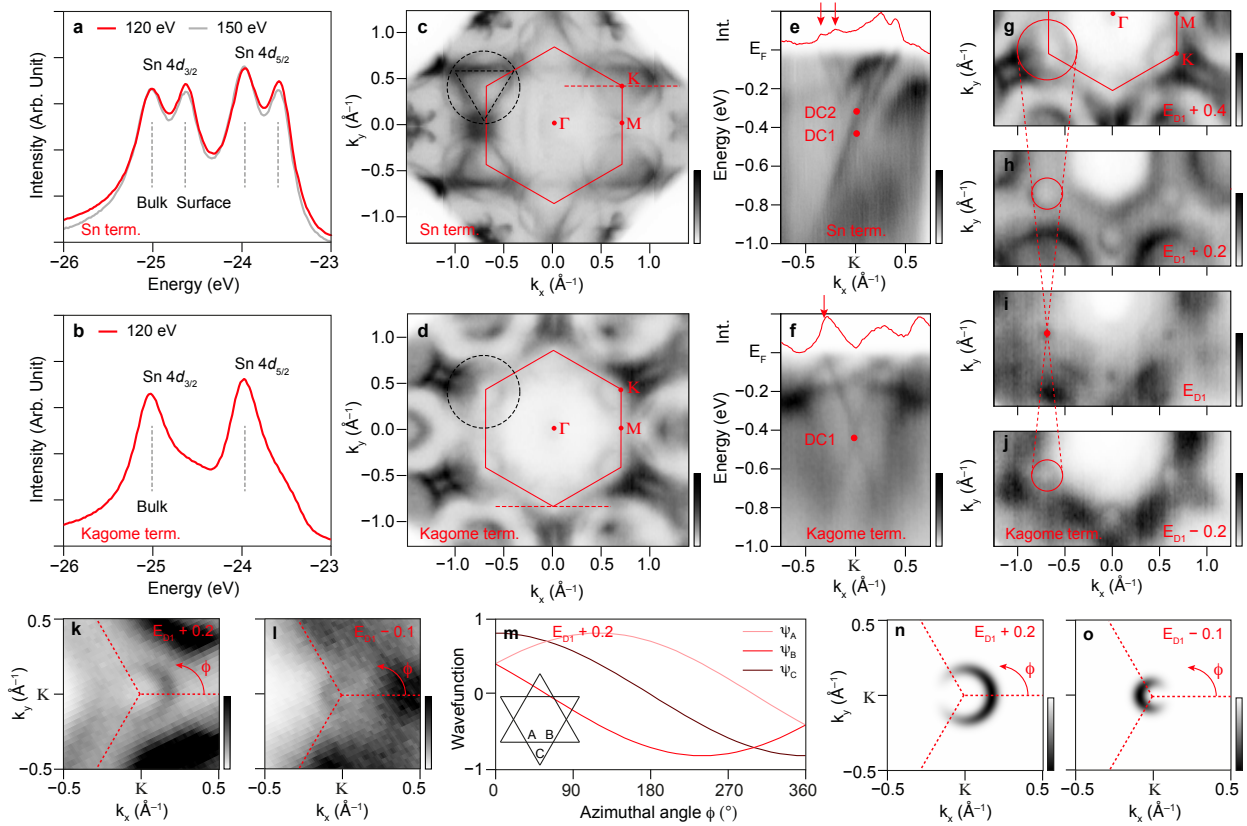

**Figure 2 | Photoemission experiments on FeSn. a,b,** Two representative XPS spectra of *in situ* cleaved FeSn, from which we identified the surface termination as Sn and kagome layers respectively. **c,d,** Fermi surfaces of FeSn measured on Sn and kagome terminations respectively. The Brillouin zone is marked with the red solid hexagon. Dashed black circle and triangle schematically represent the prominent Fermi surfaces centered at the K-point. **e,f,** Energy-momentum dispersion of FeSn across the K-point along the dashed red line marked in c,d. DC1 and DC2 indicate the position of Dirac points. Momentum distribution curves at Fermi energy are overlaided, with arrows indicating peaks at $k_F$. **g-j,** Constant energy maps measured on the kagome termination at +0.4 eV, +0.2 eV, 0 eV, and −0.2 eV with respect to the $E_{D1}$, with clear shrinking and reopening of the Dirac pocket as the energy crosses the Dirac point. **k,l,** Constant energy map above and below DC1 respectively highlighting the modulation of photoemission intensity around the Dirac point. **m,** Dirac wave function on three sublattices of kagome lattice featuring Berry phase π acquired after 2π azimuthal rotation. **n,o,** Simulation of sublattice interference pattern of kagome lattice from Dirac wave functions in m. Colorbar at the bottom right of each panel indicates intensity from minimum (bottom) to maximum (top).

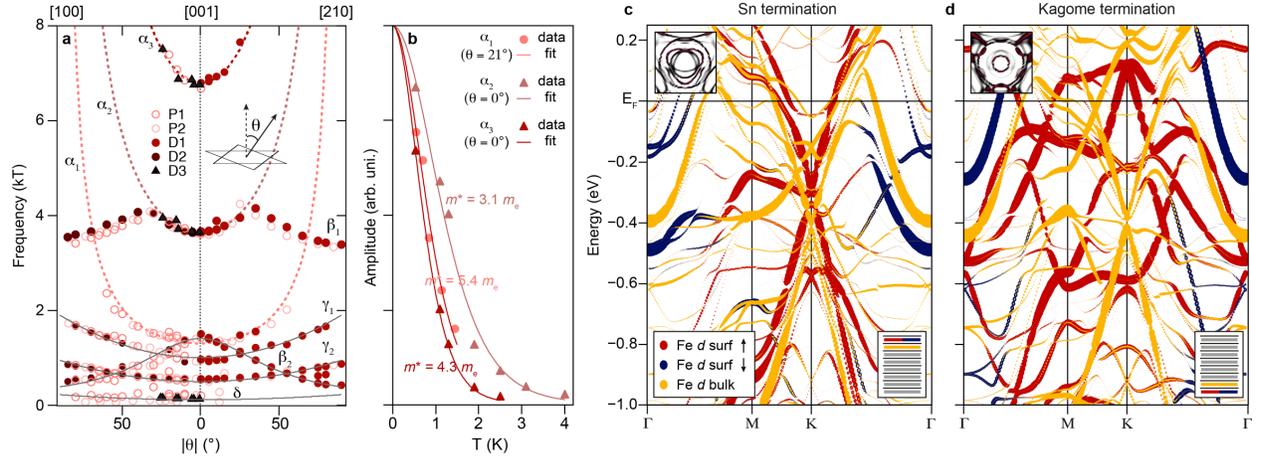

**Figure 3 | Magneto-quantum oscillations and slab DFT calculations of FeSn. a,** Angular dependence of dHvA (circles) and SdH (triangles) oscillation frequencies of FeSn. Hollow symbols indicate the frequencies obtained from pulsed field measurement, while filled symbols indicate those obtained from DC field measurement. Different colors indicate different samples. The inset schematically represents the definition of θ with respect to the kagome plane normal. Dashed lines indicate the fit of α bands to 1/cosθ function, the behavior expected from a quasi 2D Fermi surface. **b**, Temperature dependence of quantum oscillation (dHvA for $\alpha_1$, SdH for $\alpha_2$ and $\alpha_3$) amplitude of α bands. Solid lines are fit to the Lifshitz-Kosevich formula (see Methods). All curves are normalized to the expected zero-temperature value. **c,d,** Slab DFT calculations of eight FeSn structural unit cells. Bottom-right insets show schematic structure of the slab with black lines representing Sn layers and rectangles representing kagome layers. Red and blue circles represent the weight of spin up and down states on topmost kagome layers ('surf'), while orange circles represent the weight of states on the second topmost kagome layers ('bulk'). The Top-right insets in d,e show the contours of Dirac pockets around the K-point at the Fermi level, which capture the anisotropic warping of bulk (DC1) and surface (DC2) Dirac cones (see Fig. 2c,d for comparison).

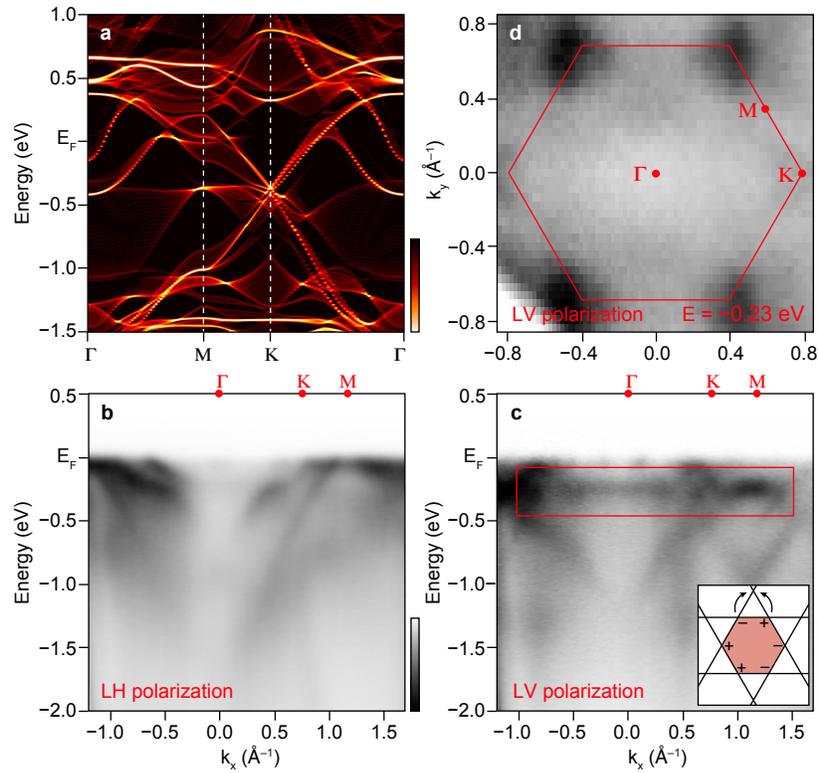

**Figure 4 | Signature of flat bands in FeSn. a,** Energy-momentum dispersion of FeSn integrated along the out-of-plane momentum ($k_z$). **b,c,** Experimental band structures of FeSn along the Γ-K-M high symmetry directions measured with LH and LV polarizations (incident photons in- and out-of-the photoelectron emission plane, respectively) on kagome termination. The red box in c highlights the nearly flat band around $E_{flat} = -0.23$ eV. Inset in c depicts the confinement of electron in hexagon of kagome lattice arising from the destructive phase interference between hoppings from different sublattices. **d,** Constant energy map of FeSn at $E_{flat}$ highlighting the almost uniform intensity distribution from the nondispersive band (the higher intensity at a few of the K-points is due to the presence of the Dirac band in the same energy range).

| ARPES | | | |
|---|---|---|---|
|  | $A_F$ (Å$^{-2}$) | $v_F$ (10$^5$ m/s) | $E_D$ (eV) |
| Dirac Cone 1 (DC1) | 0.38 ± 0.03 | 1.70 ± 0.20 | 0.43 ± 0.02 |
| Dirac Cone 2 (DC2) | 0.26 ± 0.02 | 1.87 ± 0.20 | 0.31 ± 0.02 |

| dHvA | | | | | |
|---|---|---|---|---|---|
|  | $f_0$ (T) | $m^*$ ($m_e$) | $A_F$ (Å$^{-2}$) | $v_F = \sqrt{\frac{E_{D1}}{m^*}}/\cos\theta$ (10$^5$ m/s) | $v_F = \sqrt{\frac{E_{D2}}{m^*}}/\cos\theta$ (10$^5$ m/s) |
| $\alpha_1$ | 1310 ± 2 | 5.4 ± 0.4 | 0.1251 ± 0.0002 | 1.17 ± 0.07 | 1.04 ± 0.05 |
| $\alpha_2$ | 3641.5 ± 0.1 | 3.1 ± 0.2 | 0.34761 ± 0.00001 | 1.54 ± 0.09 | 1.34 ± 0.09 |
| $\alpha_3$ | 6755.3 ± 0.8 | 4.3 ± 0.3 | 0.65586 ± 0.00008 | 1.31 ± 0.08 | 1.13 ± 0.08 |

**Table 1 | Comparison of ARPES and dHvA experiments on FeSn.** $v_F$ from ARPES is calculated from the slope of Dirac bands at the Fermi energy averaged on multiple momentum space directions to account for the trigonal warping. $v_F$ from dHvA is derived from the effective $m^*$ and $E_D$ as $v_F = \sqrt{\frac{E_{D1}}{m^*}}/\cos\theta$ assuming Dirac dispersion.